# Tracing high-profile attention to questionable research as a case for funder due diligence

Federica Silvi* — Digital Science, UK

ORCID: https://orcid.org/0009-0001-2532-7961

LinkedIn: https://www.linkedin.com/in/federica-s-6544472a/

Leslie D. McIntosh — Digital Science, UK

ORCID: https://orcid.org/0000-0002-3507-7468

LinkedIn: https://www.linkedin.com/in/leslie-mcintosh/

*Corresponding author. Email: f.silvi@digital-science.com

## Abstract

When questionable research has entered the scientific record, it stands the same chance as other literature of shaping national and international policy, clinical guidance, and research and development activities. This study uses a known authorship-for-sale network identified in August 2022 to examine whether questionable research influences policy, patents, and clinical guidelines, and whether its authors continue to secure funding and publish after exposure. Across nearly 2,000 publications in our dataset, 57 were cited in policy documents, 12 in clinical guidelines, and 480 in patents. Nearly a quarter of these publications are linked to one or more grants: funding that could otherwise have supported rigorous, ethical science. Among the 278 authors we traced, publishing and funding continued well after the PNN's public exposure in 2022: over 90% continued publishing, and 25% were linked to a grant.

As our findings show, the outputs of paper mills and authorship-for-sale networks can continue to inform policy and clinical practice and support their authors' career progression even after the questionable practices behind them are exposed. We present a case for funders to consider authorship and network structure as part of a holistic due diligence process, alongside metrics such as citation counts and attention data.

# Introduction

Scholarly output has been growing exponentially for decades, with global publication counts increasing at a sustained annual rate of around 5% since 1952 [1]. A measurable share of this expansion is now attributable to paper mills (fraudulent organisations that sell authorship positions on pre-written manuscripts [2]) and similar operations. Their activity has grown sharply since 2016 – the year the first paper mill-linked retraction was recorded [3], outpacing the growth of legitimate science itself [4]. Their influence persists even after exposure: within the life sciences the number of systematic reviews citing retracted paper mill articles rose from just one in 2015 to 105 in 2023. The trend is likely understated even in the most recent data since retraction typically lags well behind publication and citation [5].

Growing numbers of publications do not necessarily result in more knowledge the world can benefit from: in fact, the larger the volume, the greater the risk of polluting scientific literature with unverifiable results and low-quality work, practices which almost 40% of researchers report having witnessed in colleagues [6].

Paper mills alone are estimated to have advertised over 20,000 authorship positions for sale across multiple countries [7]. Whether driven by volume, or by compromised editorial gatekeeping, this pollution erodes public trust in science and opens the door to disinformation [8]. But there are more tangible consequences, too: when scarce funding supports less innovative or even fraudulent work, the system actively squanders the opportunity to invest in rigorous, ethical science [9].

The roots of that squandering lie partly in the factors that inform career advancement and funding decisions: both are driven largely by measures of visibility and output, which often clash with the pursuit of making reliable science available to the community [10, 11]. Understanding this tension means looking closely at what these measures capture and what they miss.

Citation-based indicators such as the h-index remain the dominant measure of academic impact, but are narrow by design: a paper's first organic citation can take years, and influential work can remain uncited [12, 13]. The Journal Impact Factor compounds this problem in a different way [12, 13]: a metric built to describe a journal's citation average gets applied, often incorrectly, to judge individual articles within it. These measures are also directly gameable, through practices such as citation cartels and coercive citation [14].

Altmetrics emerged in 2010 as a complementary approach to those traditional metrics [13], capturing more immediate attention across platforms such as social media, blogs, news outlets, reference managers, and increasingly policy documents, patents, and clinical guidelines [15]. Altmetrics also face a parallel concern about metrics gaming [13]; while documented cases of researchers caught deliberately inflating their own attention scores are rare, the underlying data remains vulnerable to automated distortion. For example, bot accounts generating disproportionate shares of social media mentions [16].

Institutional and funder-level uses of altmetrics, from evaluating researchers for prizes and grants to funder reporting, have been proposed or investigated in previous studies [15, 17] and, in some settings, already adopted [18, 19]. Within the broader landscape of the sources monitored by altmetrics, mentions of research in policy documents, clinical guidelines and patents occupy a distinct position [15, 20]. These mentions signal that research is entering the evidence base drawn on by national governments, intergovernmental organisations, and clinical practice leaders [21].

Unlike social media attention, which is immediate and high-volume, policy and guideline citations are rare across research in general [22], though markedly higher within research specifically produced to inform policy and clinical practice [21]. This attention typically surfaces more slowly than mentions in social media, news outlets or blogs – accruing over an average lag of almost four years after publication. Yet, these citations and references are more durable, and funders, institutions and businesses track them closely [15].

Patent citations serve a related function. Because most patent applications are made with the goal of securing or preserving commercial advantage, a citation to research within a patent signals that the research has fed into a process with real-world commercial or technological consequence [23], evidencing the kind of knowledge transfer between academic research and industry that funders and institutions also have reason to track [24]. A policy citation, patent reference, or clinical guideline mention can therefore actively raise confidence in the author's body of work, strengthening a case for continued or renewed investment [15].

Funders already carry some formal responsibility for the integrity of the work they support

[25], yet due diligence at the point of funding rarely extends to scrutinising how attention accumulated, or whether it reflects genuine influence rather than volume, coordinated activity, or exploitation of legitimate channels. Closing this gap means looking beyond the citing document itself, to the authors being cited: their body of work, their collaboration networks, and how their funding and publication activity behaves over time.

This analysis explores three research questions using a known authorship-for-sale network:

1. RQ1: Does questionable research generate attention in policy documents, patents, and clinical guidelines?
2. RQ2: Are its authors receiving legitimate grant funding?
3. RQ3: Do they continue to publish and receive funding after the network they were part of or their questionable research practices have been exposed?

# Methods

This paper presents a forensic scientometric case study of the Pharmakon Neuroscience Network (PNN), an authorship-for-sale network publicly identified in August 2022 [26, 27] and previously characterised as a collective of over 300 authors across 40 countries [28].

This Methods section summarises the study's data sources and analytical procedures. A Supplementary Information file provides full procedural detail needed to reproduce each step of the analyses, including study limitations.

## Data sources

This study combines data from Dimensions (publication metadata; researcher profiles; grant records) [29], Altmetric (attention data, principally from policy documents and clinical guidelines) [30], Policy Commons (organisational category classifications) [31], the Retraction Watch Database (primary source for the retraction status of publications) [32], and PubMed's E-utilities API (retraction status fallback) [33]. Dimensions and Altmetric are Digital Science products, accessed via employee licences; use of these tools is disclosed further in the Declaration of Interest statement.

Altmetric data was collected via the Altmetric Explorer and Details Page API on 9 April 2026; Dimensions data was collected via the DSL API (dimcli v1.7) on the same date; the Retraction Watch Database snapshot was downloaded on 14 April 2026. Mention-level date data for X (formerly Twitter) posts was additionally collected via the Explorer API on 4 August 2026 for informal contextual review; these mentions are not included in the quantitative counts reported in this study. The post-exposure author analysis was re-run on 18 August 2026 using a later DSL API client version (dimcli v1.8/DSL v2.15).

All the above data sources update continuously, so the figures reported here reflect frozen snapshots taken on the dates specified above; the underlying dataset is available on Figshare (see Data availability). Altmetric's coverage of policy and guideline sources, while extensive, is not exhaustive, so the mentions analysed here reflect what Altmetric had indexed at collection time, not a complete account of all citing activity.

Data processing and analyses were conducted in Python 3 (the default Google Colab runtime at the time of analysis), using Jupyter notebooks hosted on Google Colab.

## Dataset construction

This study draws on three data collections, corresponding to its three research questions (RQ) and reconciled by Dimensions ID where they overlap. Full details on how each dataset was compiled, including the exact spreadsheet tabs deposited alongside this paper, is available as Supplementary Information (see Data availability statement). Figure 1 summarises the scope of all datasets and the research questions each informs.

**Figure 1:**

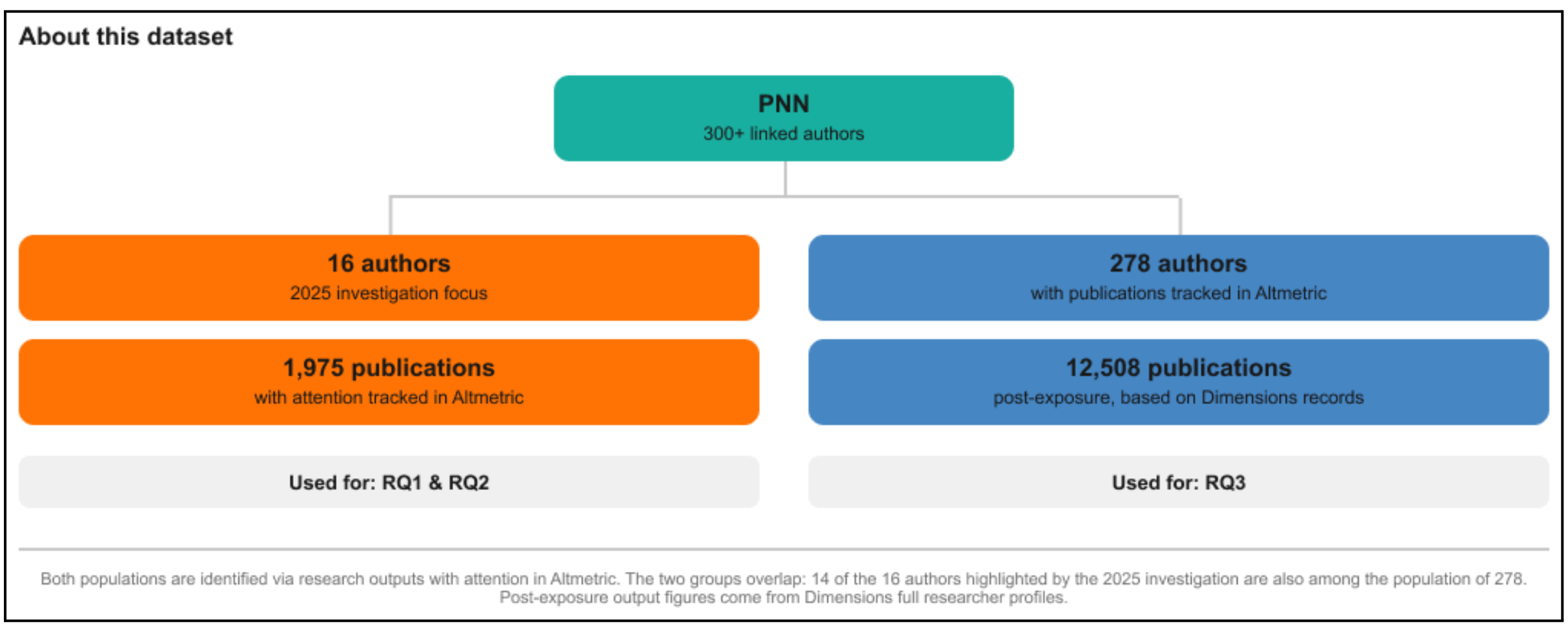

RQ1 draws on the publication record of 16 authors linked to the PNN through a 2025 investigation into its funding activity, combined with publications naming the PNN directly as an author affiliation, funder, or acknowledged entity at the time of its public exposure (restricted to publications with 25 or fewer authors, since very large author lists make individual accountability difficult to assess). The dataset focuses specifically on publications with Altmetric-tracked attention. All policy and clinical guideline mentions received by this combined population were examined (see Mention curation and verification below), and patent mentions were counted as a further indicator of attention.

RQ2 uses the same combined population as RQ1. Grant records were examined across its full body of publications, then narrowed to the subset of grants already identified by the 2025 investigation as having drawn specific scrutiny (see Grant linkage below).

RQ3 uses a list of 278 unique authors derived from a selection of research outputs linked to the PNN at the time of its public exposure. This is a broader population, used to test whether the patterns found for a smaller sample of individuals generalise across the wider network.

The PNN's public exposure date on 24 August 2022, where it was mentioned in Retraction Watch, is used throughout as the cutoff between "pre-exposure" and "post-exposure" for publications, mentions, grants, and retractions. Because of the lag between submission and publication, some publications dated shortly after this cutoff may have been submitted before the exposure.

Author names are not included in the deposited dataset (see Data availability statement). Individual researchers are instead referred to by an arbitrary “Researcher N' pseudonym” used consistently throughout this paper and the deposited dataset; the underlying analysis identifies authors by Dimensions researcher ID.

## Policy, clinical guideline and patent mention curation

Every Altmetric-tracked policy and clinical guideline mention was individually checked against the citing document's full text, where accessible, confirming the PNN-linked publication in its bibliography and, where relevant, its in-text citation. To provide additional interpretive context for the policy and guideline mentions in this study, the text surrounding each valid mention was manually assessed for sentiment, using classification criteria developed for Altmetric's programmatic sentiment analysis of social media mentions [34].

Mentions were deduplicated across datasets using a combined key of Dimensions ID, citing document URL, and mention type. 28 policy mentions that fell into the below categories were flagged for exclusion:

- The citing document was no longer accessible via its original URL at the issuing organization’s website. Policy documents that are no longer publicly accessible are retained in Altmetric's database as a historical record; however, their content can no longer be independently verified. Documents flagged as inaccessible via their original URL were additionally checked against the Internet Archive's Wayback Machine before exclusion.

- The citing document was a duplicate of another mention already counted.
- The mention was a false positive (i.e. the cited publication was not the one Altmetric had attributed, or Altmetric's automated source classification had misidentified a peer-reviewed journal article as a policy document).

All clinical guideline mentions in our dataset are included in the analysis; no duplicates or ineligible documents were identified. Every publicly accessible guideline document was checked via its URL, in the same manner as policy documents. The majority of guideline mentions in this dataset are drawn from journal articles indexed in Dimensions; where the original article was behind a publisher paywall, we have considered the list of references on the publication’s Dimensions record as confirmation that a mention exists.

Citing policy organisations were categorised (IGO – Intergovernmental Organisation; Government; NGO – Non-Governmental Organisation; Nonprofit; Think Tank; Corporation) primarily via Policy Commons, or via the organisation's own website where no Policy Commons profile existed. Guideline-issuing organisations were classified by clinical area according to each society's stated scope; societies covering multiple areas were classified by their primary listed focus. All source pages consulted are linked in the deposited dataset.

Patent mentions, reported as a quantitative measure of tracked attention, were not subject to the same manual verification, owing to their substantially higher volume, and should be read accordingly. To provide interpretive context, the text surrounding each valid policy and guideline mention was initially discussed by the authors then manually assessed for sentiment by a single reviewer (FS), adapting classification criteria built and validated for social media posts rather than policy or guideline text.

## Grant linkage

Grants were linked to publications via Dimensions' supporting_grant_ids field, capturing grants where one of the authors in this analysis may be named investigator or co-author of the grant's resulting publications. Because this method cannot always distinguish the two, this paper describes grants as authors 'held or were linked to,' rather than assuming investigator status throughout.

Because Dimensions records only a grant's start year, grants starting in 2022 cannot be placed definitively before or after the 24 August cutoff, and are conservatively classified as post-exposure.

## Post-exposure activity per author

For each author in the deposited dataset, we queried Dimensions for their complete publication and grant history, independent of PNN-named or Altmetric-tracked outputs, and logged records dated after the exposure cutoff. For grants: start years, grant URLs, and total count per author. For publications: post-exposure publication counts per author, and the Dimensions IDs of post-exposure publications.

## Retraction status determination

Retraction status of a publication was assessed in two stages: first by DOI lookup against the Retraction Watch database; then, where no record was found, through a fallback query against PubMed's E-utilities API for a “Retracted Publication” record.

Retraction status is reported using three categories: “retracted”; “not retracted; and “no retraction on record” (i.e. not found in either source). The last category reflects an absence of evidence in these two sources, not a confirmed absence of retraction. Because Retraction Watch records only retracted publications, “not retracted” statuses were confirmed only via PubMed.

## Statistical approach

This study reports descriptive counts and proportions; no inferential statistical tests were applied, and no claims of statistical significance are made.

# Results

## Questionable research and real-world attention

RQ1: Does questionable research generate attention in policy documents, patents, and clinical guidelines? We developed a set of 1,975 publications with Altmetric-tracked attention; within this set, 57 publications received at least one policy mention (76 mentions in total from 27 organisations), 12 received at least one clinical guideline mention (22 mentions from 18 organisations), and 480 publications were mentioned in patents (2,623 patent mentions in total). (Table 1)

**Table 1 | Policy, clinical guideline, and patent mentions obtained by publications in our dataset (n = 1,975; source: Altmetric Explorer, 9 April 2026)**

| Mention type | Mentions | Publications | Organisations |
|---|---|---|---|
| Policy | 76 | 57 | 27 |
| Clinical guidelines | 22 | 12 | 18 |
| Patents | 2,623 | 480 | N/A |

Policy citations were concentrated among intergovernmental and government organisations rather than smaller or commercial sources (Fig. 2). The largest single source was the Food and Agriculture Organization of the United Nations (FAO), followed by the Publications Office of the European Union, the Organisation for Economic Co-operation and Development (OECD), and the Canadian Agency for Drugs and Technologies in Health (CADTH) (Fig. 3). A number of these mentions reflect the same organisation citing the same author's work more than once, or citing several different publications by one author. The case studies described below examine this pattern in more detail.

**Figure 2:**

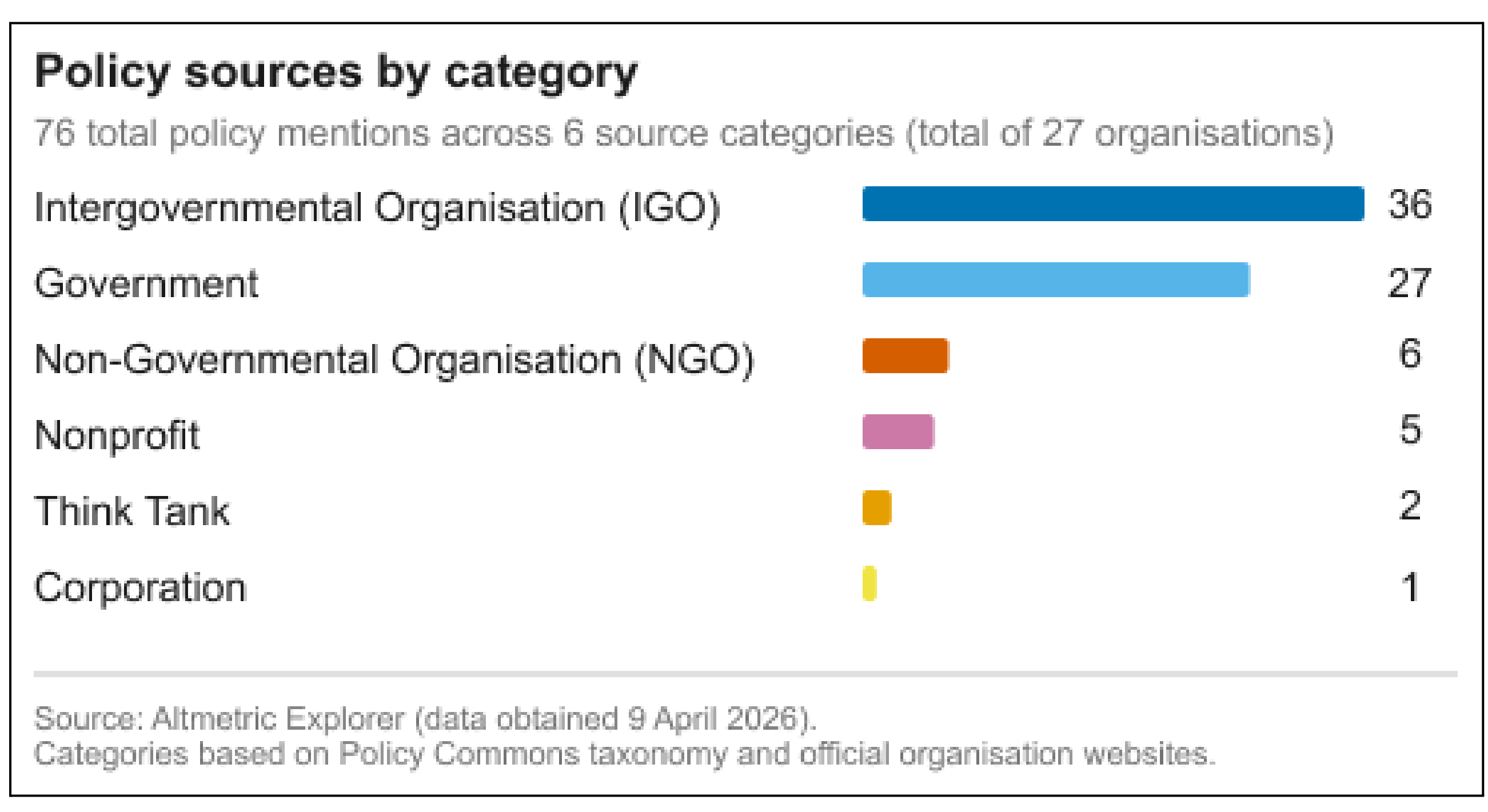

**Figure 3:**

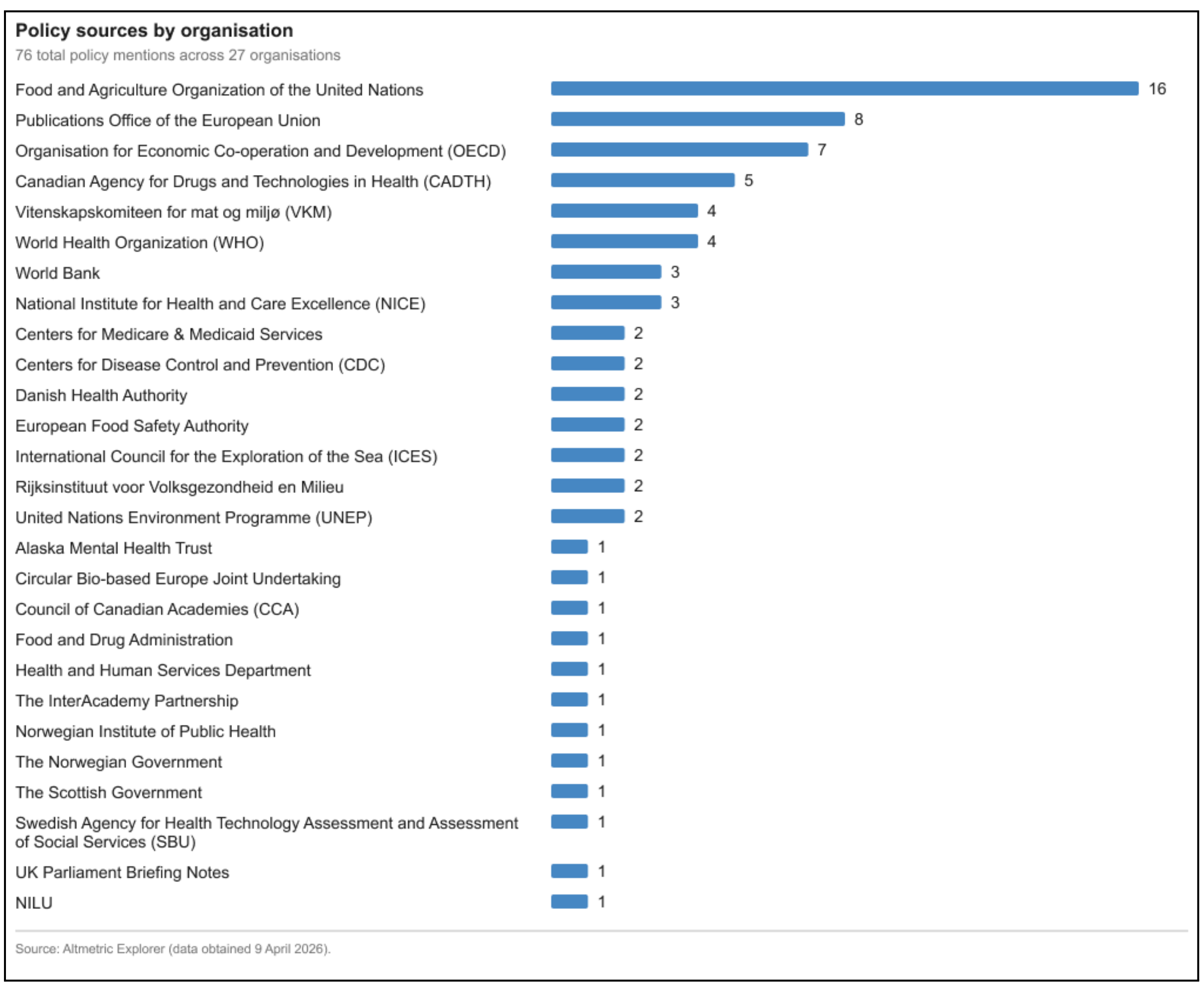

Clinical guideline citations were more dispersed than policy citations, spanning internationally recognised medical societies including the European Society of Cardiology, the American College of Chest Physicians, and the National Comprehensive Cancer Network, alongside national and specialty societies in haematology, neurology, vascular medicine, endocrinology, rheumatology, gynaecology, and veterinary emergency medicine (among others). As defined in Methods, clinical guidelines offer direct recommendations to practising clinicians, distinct from the broader advisory role of policy documents; a citation in a guideline therefore signals entry into a specific area of practice recommendation, regardless of how many other mentions accompany it. By this measure, research by PNN-linked authors reached practice recommendations across 13 distinct clinical areas.

## Attention in practice: three researcher case studies

Once questionable research is published, it can accumulate attention that can be used to build a compelling-looking funding narrative. Three cases from our dataset illustrate what that attention looks like in practice.

**Case study 1**: Researcher 244 accounts for 18 of the 22 clinical guideline mentions in our analysis. Twelve of these reference a single 2013 publication, itself a clinical guideline, cited by major medical associations as recently as 2024. In the same year, two separate US government documents by the US Centers for Medicare & Medicaid Services (CMS) and the US Health and Human Services (HHS) Department mentioned a different grant-funded paper by Researcher 244. Both in-text citations highlight that the US Office of Research Integrity (ORI) found the paper to be involved in a research misconduct case. Retraction Watch also covered the case [35], and the paper was since retracted.

Beyond these two papers, Researcher 244 has a further 15 works mentioned 23 more times in policy and clinical guidelines, none retracted to date. Researcher 244 was linked to six grants between 2013 (the problematic paper's publication year) and 2024 (when the misconduct case was reported).

This case study shows that a misconduct ruling may carry no real cost to a researcher beyond the specific publications it concerns, even when confirmed by authorities such as a federal agency.

**Case study 2**: Researcher 105 has 29 publications mentioned 37 times across 33 unique policy documents and one clinical guideline. Their publication record stretches back to 1990, but 53% of their output appeared in just four years (2020–2023) alongside a co-author network of 1,968 people: a pattern consistent with participation in an authorship-for-sale scheme [27].

Researcher 105 co-authored a paper cited in a 2019 European Commission technical report on fertilising materials, which could form the legal basis for EU regulation on waste-derived fertilizers. The paper is cited in the following passage [36]:

“The database for this class of compounds is too weak to support that statement...Rey-Salgueiro et al. (2016) also indicated that the concentrations of benzene, toluene, ethylbenzene, the ortho-, para- and meta-xylenes and styrene (BTEX + S) in all samples analysed in their study were low for bottom and fly ashes with maximum concentrations of 0.3 mg kg-1.”

The paper itself has not been retracted, and its scientific content is not disputed here. However, its presence in a document shaping EU food safety policy illustrates how questionable affiliations can reach into high-stakes policy contexts without any integrity flags being raised.

**Case study 3**: A Canadian Agency for Drugs and Technologies in Health (CADTH) policy document cites a single paper co-authored by three of the researchers examined in this study. Two of these are among the researchers whose funding activity was investigated in 2025; the third is the author whose publication activity first prompted the PNN network's public exposure. The paper is cited in the following passage [37]:

“First-line treatments for mild to moderate AD are the cholinesterase inhibitors (ChEIs) donepezil, galantamine, and rivastigmine, which increase concentrations of neuronal acetylcholine and improve symptoms such as memory loss and cognition.”

As with Researcher 105's case, the paper itself has not been retracted, and its scientific content is not disputed here. This case is presented to show that attention converging on a single publication can extend across several members of the same network at once, adding to the profile of more than one author simultaneously.

## Grant funding linked to questionable research

RQ2: Are the authors of questionable research receiving legitimate grant funding? A prior investigation into the Pharmakon Neuroscience Network's funding activity found that more than 20 authors linked to the network held legitimate grants worth over $15 million (US equivalent) since the network's exposure in 2022 [9].

Of the 1,975 publications tracked in the dataset described in Methods, 450 (22.8%) have at least one grant linked in Dimensions. Of those publications, 129 are directly linked to grants the 2025 investigation (by LD McIntosh) [9] focused on – 28.7% of all grant-funded publications in our dataset. Within these 129 publications, attention was concentrated in patents rather than policy documents (Fig. 4). All of the policy-cited publications and 41 of the 43 patent-cited publications pre-date the exposure of the PNN, showing that the attention that could strengthen future funding applications was already accumulating while the network operated.

**Figure 4:**

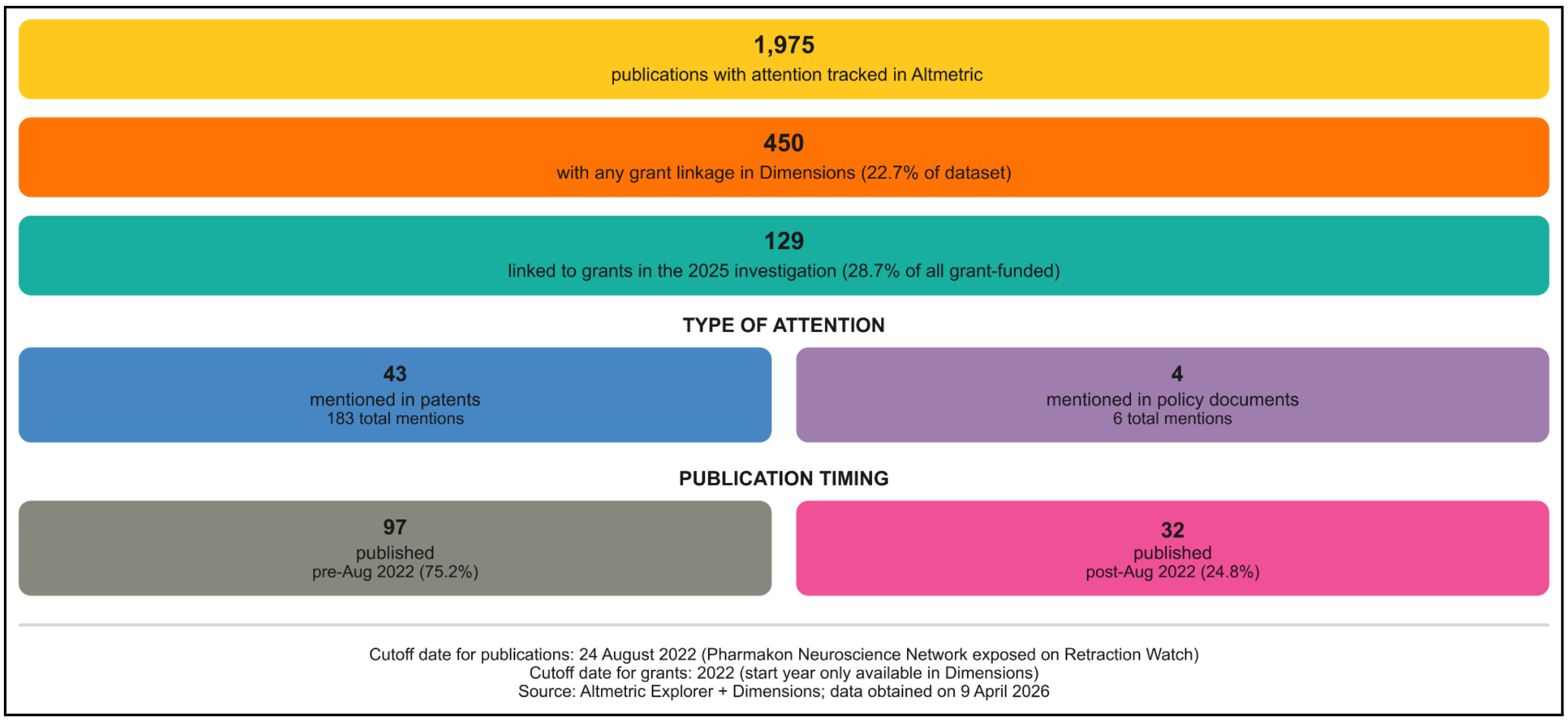

These publications mark the point at which the risk that scarce funding may be allocated to support questionable work becomes concrete: as prior work on the network has argued, "fraudulent publications can act as stepping stones to taxpayer-funded research" [9]. Here, that stepping stone is traceable to specific publications, specific grants, and specific funders.

The findings above describe scale; we now turn to what continued funding looks like at the level of an individual career by returning to Researcher 244. Researcher 244's 2013 publication reported for research misconduct in 2024, and subsequently retracted, was funded by a grant from the U.S. National Cancer Institute – a major US federal research funder. Between the paper's publication in 2013 and the misconduct report in 2024, Researcher 244 was linked to six grants in total: funding that continued not only after the Pharmakon network's exposure in 2022, but throughout the entire period the questionable 2013 paper sat uncorrected in the literature. Whether this pattern of continued funding holds more broadly across the authors examined in this study is the subject of the next section.

## Publishing and funding after exposure

RQ3: Does the output of the authors of questionable research continue after the network they were part of or their questionable research practices have been exposed? To analyse this, we drew on a third dataset described in Methods: the unique co-authors of the 98 Altmetric-tracked "Pharmakon papers" publications, comprising 278 authors in total. We traced each of these 278 authors' subsequent publication and grant activity through their complete publication records.

Although the Pharmakon Neuroscience Network was publicly identified on Retraction Watch on 24 August 2022, many of the authors involved have continued to publish since that time. These 278 authors are credited on a combined total of 12,508 papers published after the network's exposure (Fig. 5). Only 22 authors (7.9%) published nothing after that date. At the other extreme, 31 authors (11.1%) each published over 100 papers post-exposure; one author alone accounts for 567.

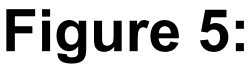

**Figure 5:**

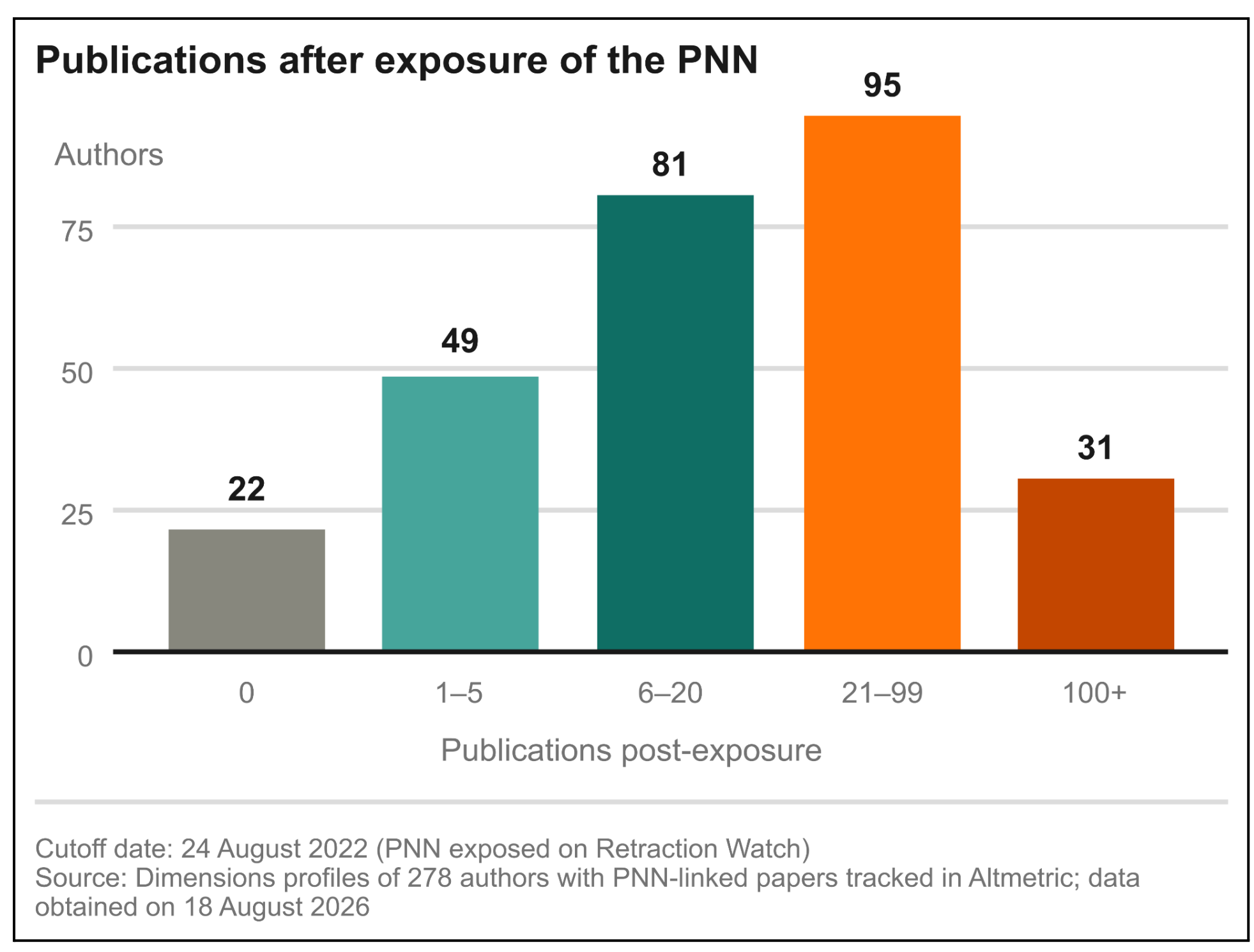

The previous subsection of Results asked whether the pattern of continued funding shown for a single author held more broadly across the wider network. The data here suggests it does: of the 278 authors we analysed, 70 (25.1%) received or were linked to at least one grant after exposure (Fig. 6), and none of these 70 stopped publishing. Three authors are linked to 10 or more post-exposure grants; separately, of the 129 publications linked to the 54 grants identified within the 2025 investigation (including grants pre-dating exposure), 32 (24.8%) were published after the network was identified (Fig. 4). Of these 54 grants, 9 (16.7%) have a start date after the network's exposure: new awards made once the PNN was already public, rather than continuing funding that pre-dated it. As previously noted, Dimensions records only grant start year, not a full date; this figure should be read as a conservative upper bound on genuinely new post-exposure awards, rather than an exact count.

**Figure 6:**

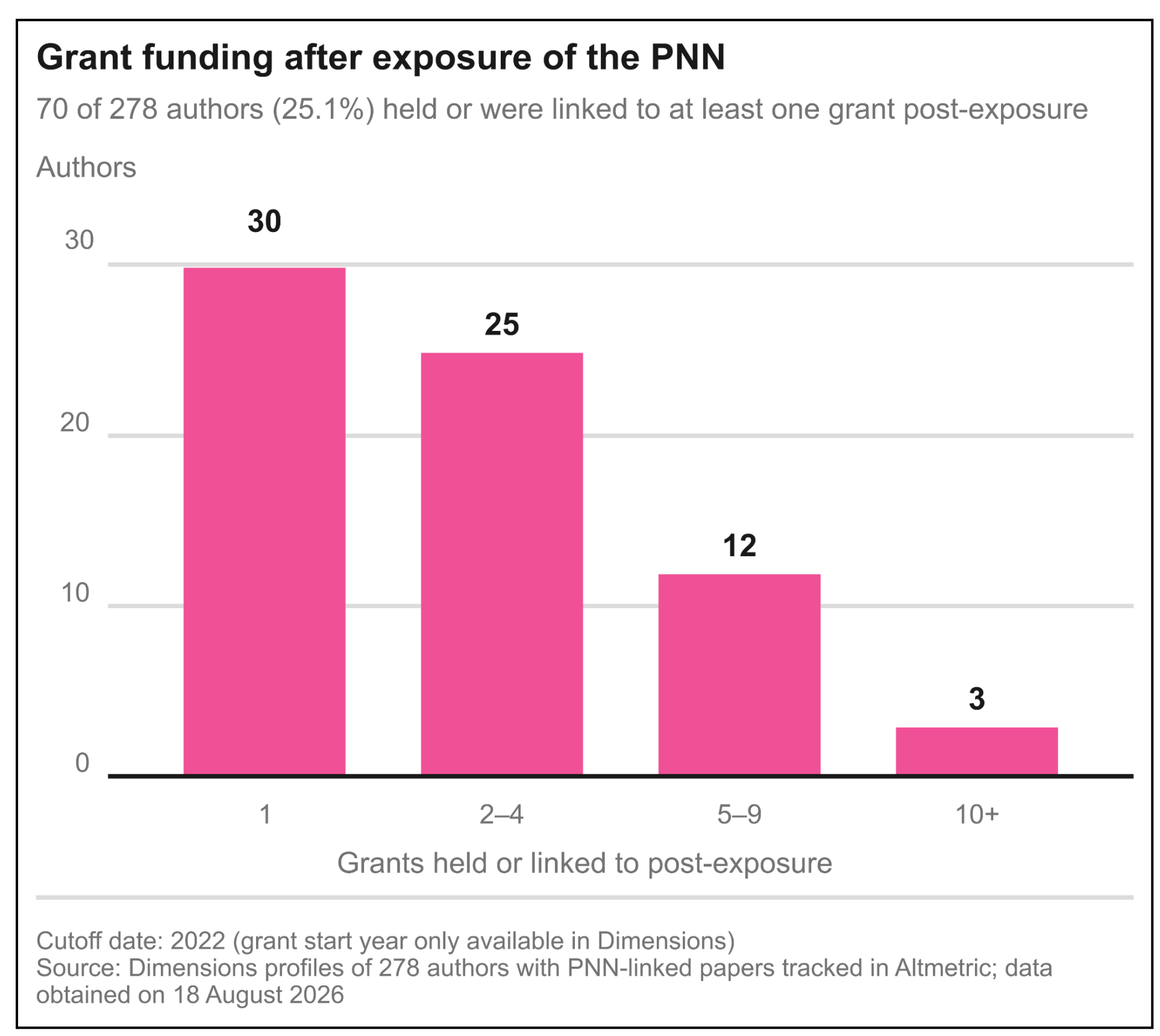

Public identification alone was also not sufficient to prompt correction of the literature. Of the 98 Altmetric-tracked publications in the "Pharmakon papers" dataset only one was retracted after exposure. The same pattern holds at the scale of the full dataset: of the 66 papers cited in policy documents or clinical guidelines across the total 1,975 publications we analysed, only one has been retracted.

This section shows that PNN-linked research reached policy documents, patents, and clinical guidelines; that its authors held legitimate public grants; and that neither pattern was meaningfully disrupted once the network became public. This points to a structural gap: exposure of a network does not, by itself, interrupt the careers or funding of its members. What responsibility that gap places on funders, institutions, and publishers is the subject of the discussion that follows.

# Discussion

Paper mill and authorship-for-sale operations are not a matter of individual researchers cutting corners: they are organised enterprises [4], differing in scale from individual academic misconduct, with many thousands of fraudulent manuscripts successfully published in peer-reviewed journals [2]. Their scale is itself a source of systemic risk to the research ecosystem: paper mill activity is now estimated to account for more than 1.5% of the research literature [2], which, at current global publication volumes, represents tens of thousands of papers [38]. Editors, policymakers, and guideline authors who encounter such work are not always positioned to recognise questionable practices. Markers of unreliable research are not always apparent [39], and detection of paper mill activity is fragmented across under-resourced, competing publishers, each seeing only part of the pattern [2, 40]. Even the peer review process can be compromised by the same networks it is meant to catch [40].

Funders occupy a distinct position in this chain of research integrity responsibility. National evaluation systems already treat evidence of societal impact as a formal part of research assessment: the UK Research Excellence Framework and Australia's Excellence in Research for Australia both do so explicitly [22]. However, the systems in place to detect research misconduct have not kept pace: policy, clinical guideline, and patent attention is often relied upon with insufficient checks on whether the underlying research, or the legitimacy of its authorship, holds up to scrutiny.

Efforts to date have concentrated on the point of submission and publication. Journals are encouraged to build and share tools that screen for submission anomalies, duplicate manuscripts, and suspicious citation patterns [2, 41] and institutions are encouraged to use scientometric data to flag collaborations that fall outside a researcher's usual network [27]. Funders are only beginning to be drawn into this picture; recent work has called for data-driven collaboration between funders, institutions, and publishers as a practice still to be built [42].

Funders already have one verification mechanism built into the publication record itself. A grant identifier in a paper's acknowledgements links the funder to the output, and, in principle, exposes the funder to reputational risk should that work later prove unreliable or fraudulent. Unless grant and ethics identifiers are consistently recorded in structured, auditable form, neither funders nor institutions can verify which of their acknowledged outputs warrant closer scrutiny [27].

Guidance for closing this gap already exists. Funders are encouraged to require explicit commitment to research integrity from those they fund, to build integrity clauses into grant contracts, and to ensure that breaches are reported promptly rather than resolved without their knowledge [25]. Comparable screening at the point of application is already implemented elsewhere: Canada's NSERC Alliance Grants programme refers flagged proposals to national security agencies as part of a standing risk assessment process [43]. This is a different risk to the one examined here, but evidence that grant-stage screening of this kind is not a hypothetical burden. What makes this system of checks worth building for research integrity specifically is the stake funders carry, and that publishers and institutions do not share in the same way. Public and philanthropic funding is finite: a grant spent on fraudulent or unreliable work, or awarded to a researcher already implicated in producing such work, displaces one that could have supported other research. A funder's own acknowledgement on a paper, meanwhile, ties its name to work that may later prove fraudulent.

Our case studies on the Pharmakon Neuroscience Network's activity offer one visible instance of the reputational and financial risk described above. This is a single documented network; our paper focuses on in-depth analysis of a small fraction of what an enterprise at this scale is capable of producing, rather than a representative sample of paper-mill activity. Also, only one publisher was directly notified by LDM of problematic publishing practices of one author. The exposure of a network through multiple reporting avenues (e.g., Retraction Watch) does not mean a publisher would be aware of the reporting or the questionable practices of their authors. The fuller scope and limitations of this study, including Altmetric's data coverage, are set out in Methods and Supplementary Information.

Nonetheless, this research shows that PNN-linked research had already been informing policy documents, patents, and clinical guidelines for years before the PNN's exposure; that its authors had already been benefiting from legitimate public grants; and that neither pattern was disrupted once the PNN's activity became public. Exposure through retractions and misconduct findings, in other words, does not reliably restrict access to funding and attention – both can continue to flow with no sign of public exposure triggering any corrective response.

Part of the reason lies in the evaluation tools funders currently tend to rely on. Citation-based indicators carry documented limitations of their own: slow to accrue, often misapplied, and open to manipulation. Attention metrics offer timely insight into research dissemination outside academia, but their primary function is to measure how much attention research receives, not whether that attention is warranted. A publication can earn a policy or clinical guideline citation while one or more of its authors obtained their place on it through a paid authorship arrangement. Citation volume or prestige alone cannot settle the question of research integrity.

Treating grant history, citation counts, or policy uptake as evidence of standing, without also examining who else appears in an author's network and how that network behaved once scrutinised, risks funding the very practices funders have a mandate to guard against. Attention metrics alone cannot resolve this, but shifting the object of scrutiny from whether attention was received to what stands behind it is a check funders are positioned to make. The evidence gathered here indicates that this scrutiny is a viable addition to existing due diligence processes.

# Author contributions

Federica Silvi, Digital Science, UK (corresponding author): Conceptualization, Methodology, Formal analysis, Investigation, Data curation, Software, Visualization, Writing – original draft

Leslie D. McIntosh, Digital Science, UK: Conceptualization, Resources, Writing – review & editing

# Declaration of generative AI use

For coding and data exploration, Claude was used by author FS over the course of the project. Multiple versions were used as the tool was updated during this period; the most recent, Claude (version 1.26832.0 (056ee2)), is the version recorded. The resulting code was independently validated by Sam Scribner (Senior Director, Enterprise Solutions & Delivery at Digital Science). Claude was also used to assist with language editing of the manuscript text. The authors confirm that the originality and accuracy of all AI-assisted content has been verified, and that the terms of use for Claude have been reviewed, confirming suitability for publication. The authors take full responsibility for the integrity of the manuscript in its entirety, including the accuracy of all references.

# Data availability statement

The dataset analysed in this study, comprising curated Altmetric mention-level data, Dimensions publication and grant metadata, and retraction-status determinations (see Methods), together with a supplementary information document providing full methodological detail, is available at Figshare ([https://doi.org/10.6084/m9.figshare.33153377](https://doi.org/10.6084/m9.figshare.33153377)) [44].

Author names have been replaced with anonymised labels in the deposited version; the identifiers required to reproduce the analysis (Dimensions IDs) are retained. This is a de-identification measure rather than true anonymisation because Dimensions IDs resolve to public researcher profiles, individual researchers remain identifiable to anyone willing to cross-reference the two. The full dataset, including author names, is the version against which the accompanying code was developed and tested, and is available from the corresponding author upon reasonable request.

The Python code used to construct the datasets and produce the results reported in this study is available at Figshare ([https://doi.org/10.6084/m9.figshare.33329838](https://doi.org/10.6084/m9.figshare.33329838)) [45].

The Retraction Watch Database snapshot analysed in this study (downloaded 14 April 2026)

is a third-party dataset available at retractiondatabase.org, subject to Retraction Watch's own terms of use [32].

# Data usage statement

The data underlying this study was obtained from Dimensions, Altmetric, Policy Commons, the Retraction Watch Database, and PubMed, as detailed in the Methods and Supplementary Information. The authors confirm that data from each source was accessed and used in accordance with each platform's terms of use and data usage policies.

# Funding

No funding was obtained for the research reported in this article.

# Declaration of interest statement

FS and LDM are employees of Digital Science. Altmetric and Dimensions, the two data platforms this study relies on, are commercial products of Digital Science, and were accessed via the authors' employee licenses rather than through a research grant or external subscription. Digital Science had no role in the study's design, data interpretation, or the decision to publish, and this research was conducted independently of any commercial objective.

LDM is acting as VeriMe’s Pre-Launch Advisor on a volunteer, unpaid basis to provide subject matter expertise, context, perspective, advice, and access to networks related to their products and services.

# Figure captions

**Figure 1.** Diagram summarising the two datasets used to trace attention and funding activity linked to the Pharmakon Neuroscience Network (PNN, 300+ linked authors): the “2025 investigation authors” dataset (16 authors, 1,975 publications with Altmetric-tracked attention, used for Research Questions 1 and 2), and the co-author population derived from the Altmetric-tracked “Pharmakon papers” dataset (278 authors, 12,508 post-exposure publications recorded in Dimensions, used for Research Question 3). Both author populations are identified via research outputs with attention tracked in Altmetric; post-exposure publication counts are drawn from Dimensions' full researcher profiles. See Methods for full dataset construction details.

**Figure 2.** Policy attention tracked for the 1,975 publications in our dataset, grouped by source category. Bar chart showing the number of policy mentions attributed to each of six source-organisation categories, totalling 76 total mentions across 27 individual organisations. Categories are assigned using the Policy Commons taxonomy, supplemented by official organisation websites where not covered in Policy Commons (see Methods). Data obtained from Altmetric Explorer on 9 April 2026.

**Figure 3.** Individual organisations citing the publications in our dataset in policy documents. Bar chart showing the number of policy mentions tracked from each of the 27 organisations that cited one or more of these publications, ranked in descending order, out of 76 total mentions. Source-category breakdown for these organisations is shown in Fig. 1. Data obtained from Altmetric Explorer on 9 April 2026.

**Figure 4.** Grant linkage and downstream attention within our publication dataset. Funnel chart showing the number of publications retained at each stage of grant-linkage identification: from the full set of 1,975 publications with Altmetric-tracked attention, to those with any grant linkage recorded in Dimensions, to those linked to a grant connected to a prior investigation in 2025 (see Methods). Grant-linked publications are further broken down by attention type (patents or policy documents, with total mention counts shown), and by publication timing relative to the PNN's public exposure on 24 August 2022. Data obtained from Altmetric Explorer and Dimensions on 9 April 2026.

**Figure 5.** Distribution of post-exposure publication counts among Pharmakon Neuroscience Network (PNN)-linked authors. Histogram showing the number of authors, among the 278 PNN-linked authors tracked in this study, falling into each band of publication counts recorded after the network's public exposure on 24 August 2022. Data obtained from Dimensions on 18 August 2026.

**Figure 6.** Distribution of post-exposure grant counts among Pharmakon Neuroscience Network (PNN)-linked authors. Histogram showing the 70 (of 278) PNN-linked authors who held or were linked to at least one grant after the network's public exposure, falling into each band of grant counts recorded post-exposure. Grant counts are based on Dimensions' recorded grant start year only (see Methods). Data obtained from Dimensions on 18 August 2026.

## About this dataset

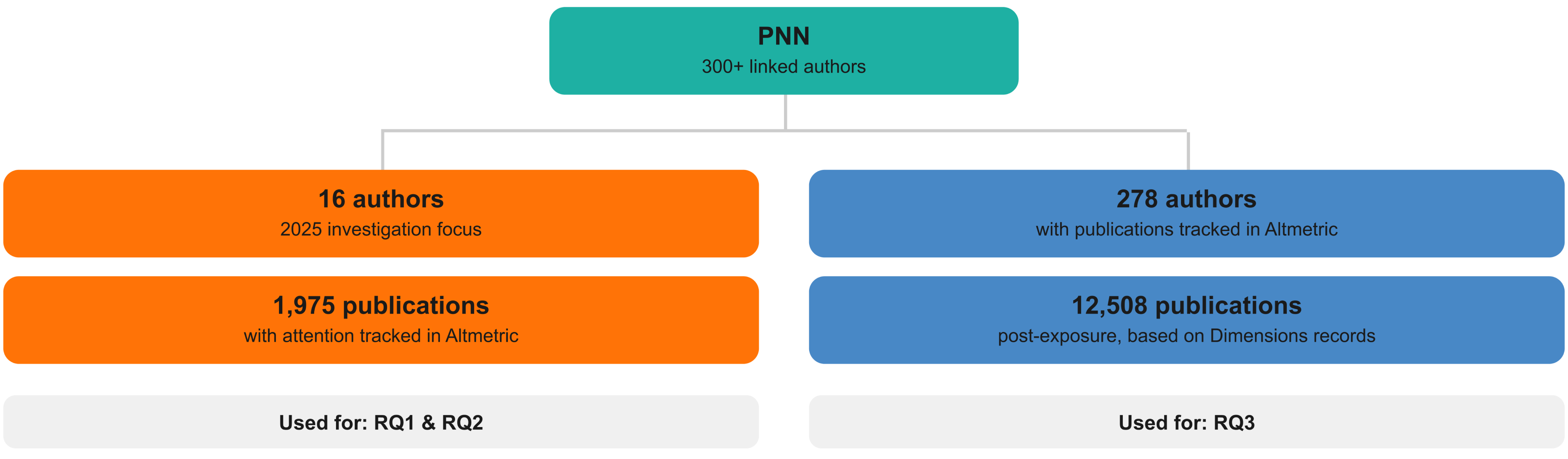

Both populations are identified via research outputs with attention in Altmetric. The two groups overlap: 14 of the 16 authors highlighted by the 2025 investigation are also among the population of 278. Post-exposure output figures come from Dimensions full researcher profiles.

## Policy sources by category

76 total policy mentions across 6 source categories (total of 27 organisations)

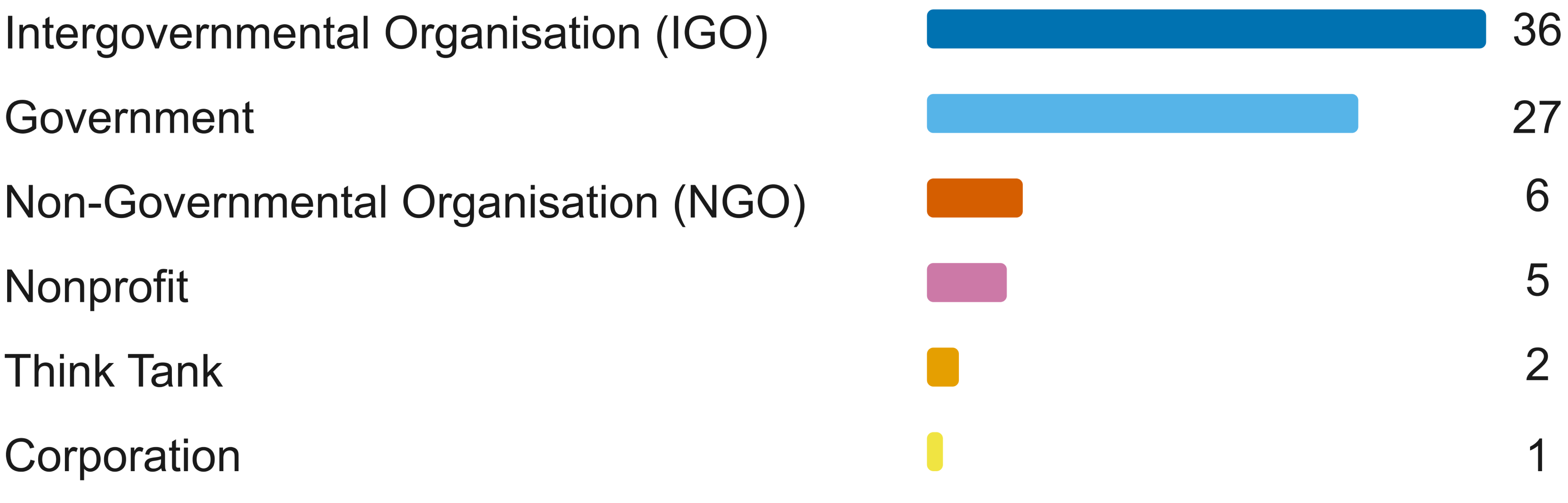

Source: Altmetric Explorer (data obtained 9 April 2026).
Categories based on Policy Commons taxonomy and official organisation websites.

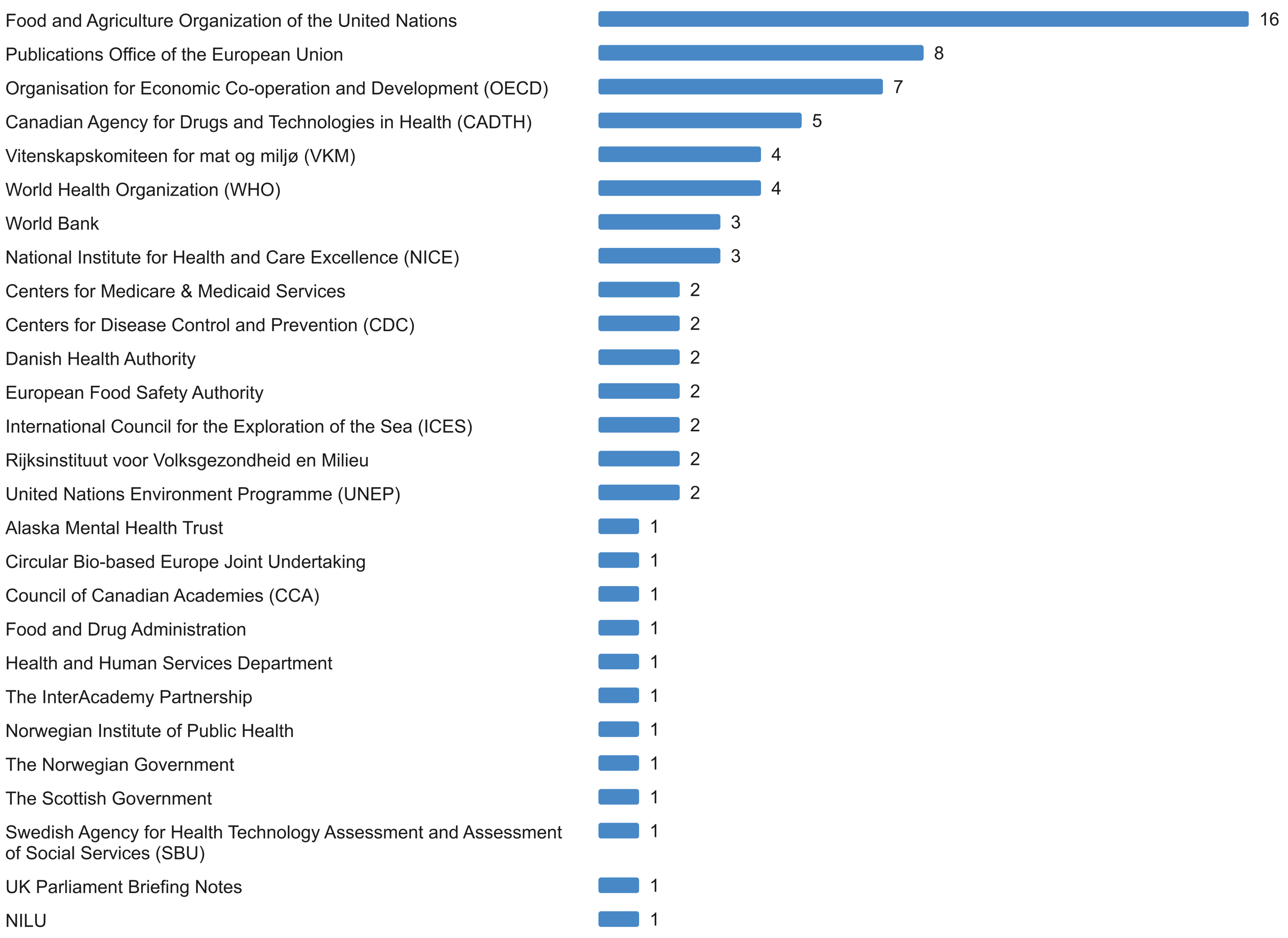

Source: Altmetric Explorer (data obtained 9 April 2026).

**1,975**
publications with attention tracked in Altmetric

**450**
with any grant linkage in Dimensions (22.7% of dataset)

**129**
linked to grants in the 2025 investigation (28.7% of all grant-funded)

## TYPE OF ATTENTION

**43**
mentioned in patents
183 total mentions

**4**
mentioned in policy documents
6 total mentions

## PUBLICATION TIMING

**97**
published
pre-Aug 2022 (75.2%)

**32**
published
post-Aug 2022 (24.8%)

---

Cutoff date for publications: 24 August 2022 (Pharmakon Neuroscience Network exposed on Retraction Watch)
Cutoff date for grants: 2022 (start year only available in Dimensions)
Source: Altmetric Explorer + Dimensions; data obtained on 9 April 2026

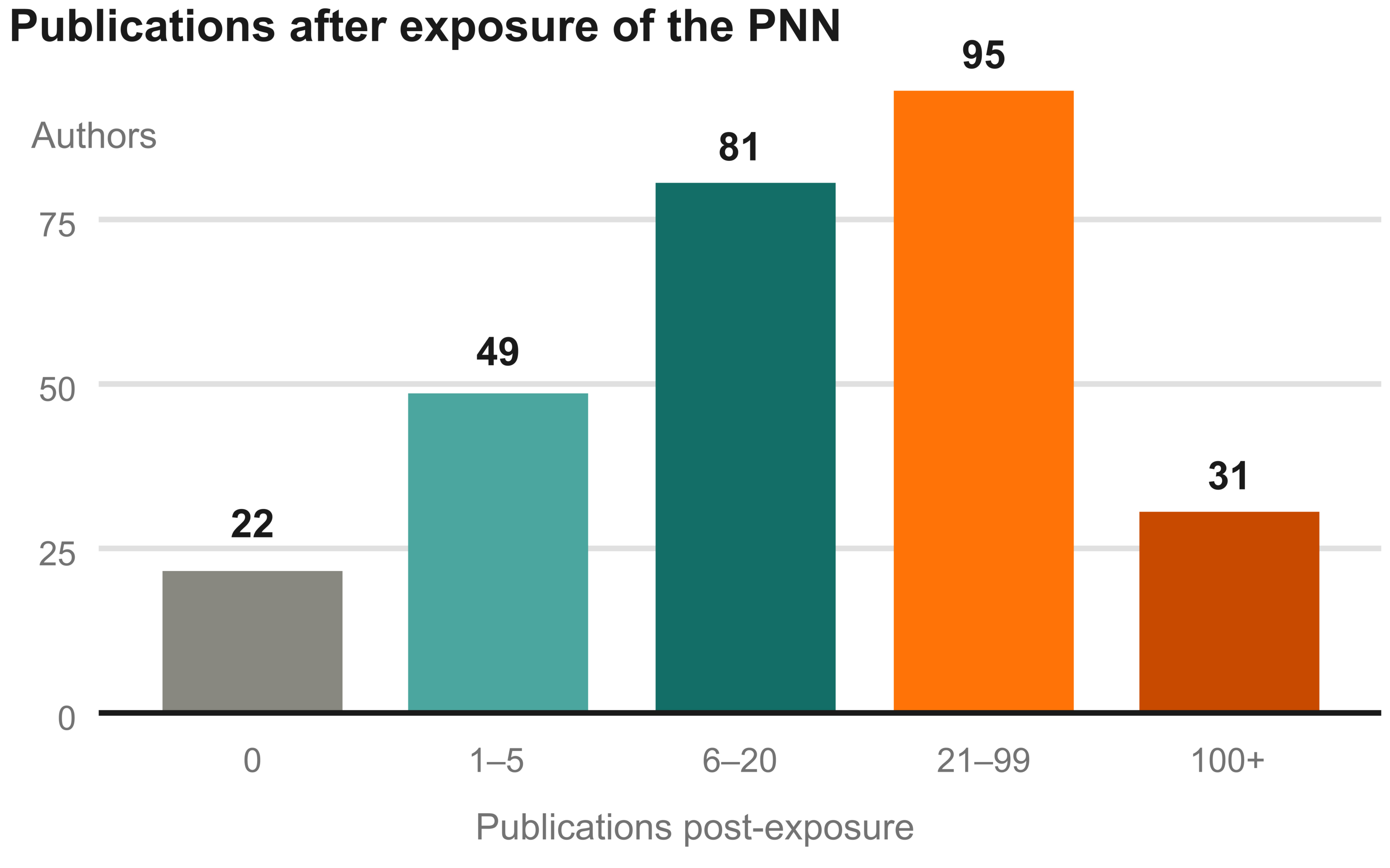

Cutoff date: 24 August 2022 (PNN exposed on Retraction Watch)
Source: Dimensions profiles of 278 authors with PNN-linked papers tracked in Altmetric; data obtained on 18 August 2026

## Grant funding after exposure of the PNN

70 of 278 authors (25.1%) held or were linked to at least one grant post-exposure

Authors

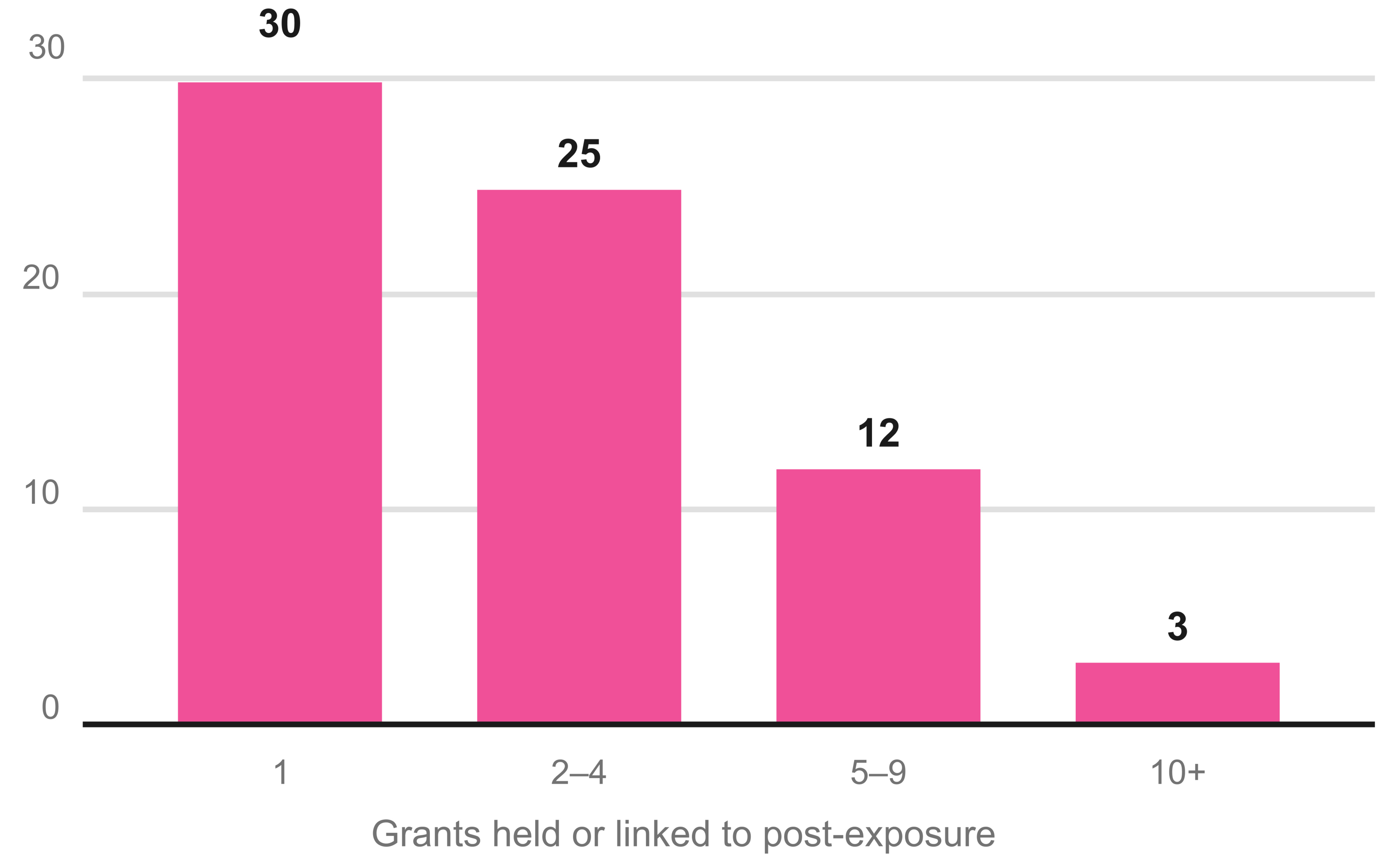

Cutoff date: 2022 (grant start year only available in Dimensions)
Source: Dimensions profiles of 278 authors with PNN-linked papers tracked in Altmetric; data obtained on 18 August 2026